\documentclass[useAMS,usenatbib]{mn2e}
\usepackage{graphics,rotating,multirow,bm,color,hyperref}

\hypersetup{
    colorlinks=true,
    linkcolor=green,
    citecolor=blue,
}

\title[Voids in Coupled Scalar Field Cosmology]
  {Voids in Coupled Scalar Field Cosmology}
\author[Baojiu Li]
  {Baojiu~Li$^{1,2,}$\thanks{E-mail: b.li@damtp.cam.ac.uk}\\
  $^1$DAMTP, Centre for Mathematical Sciences, University of Cambridge, Cambridge CB3 0WA, UK\\
  $^2$Kavli Institute for Cosmology Cambridge, Madingley Road, Cambridge CB3 0HA, UK}
\date{\today}

\pagerange{\pageref{firstpage}--\pageref{lastpage}} \pubyear{2010}
\def\LaTeX{L\kern-.36em\raise.3ex\hbox{a}\kern-.15em
    T\kern-.1667em\lower.7ex\hbox{E}\kern-.125emX}

\begin{document}

\label{firstpage}

\maketitle

\begin{abstract}
We study the properties of voids in two different types of coupled scalar field theories. Due to the fifth force produced by the scalar field coupling, the matter particles feel stronger attraction amongst each other and cluster more quickly than they do in the standard $\Lambda$CDM model. Consequently voids in the coupled scalar field theories start to develop earlier and end up bigger, which is confirmed by our numerical simulations. We find that a significantly larger portion of the whole space is under-densed in the coupled scalar field theories and there are more voids whose sizes exceed given thresholds. This is more prominent in early times because at later times the under-dense regions have already been evacuated in coupled scalar field theories and there is time for the $\Lambda$CDM model to catch up. The coupled scalar field theories also predict a sharper transition between voids and high density regions. All in all, the qualitative behaviour is different not only from the $\Lambda$CDM result, but also amongst specific coupled scalar field models, making voids a potential candidate to test alternative ideas about the cosmic structure formation.
\end{abstract}


\section{Introduction}

One of the most active research areas in modern cosmology is about the theories involving cosmological scalar fields. As a potential candidate for dark energy \citep{cst2006}, for example, a canonic scalar field could have interesting properties governed by its potential. Such quintessence \citep{wcos2000} models are studied in depth in the literature, seeing various potentials proposed, their properties investigated and their specific forms tried to be connected to developments in high energy physics. Even richer phenomenology is achieved by considering variants of the simple quintessence model, such as giving the scalar field a non-canonical kinetic term \citep{ams2000} or coupling it to the matter fields \citep{a2000} or even the spacetime curvature \citep{pb1999}, the latter case also covering certain modified gravity theories, such as the metric \citep{cddett2005} and Palatini \citep{v2003} $f(R)$ gravity and Brans-Dicke theory \citep{bd1961}. Often in these models, the behaviour of the scalar field is additionally controlled by a coupling function, and extra (fifth) forces are not uncommon. Local experiments then force the model builders to design specific mechanisms by virtue of which the (usually) severe experimental and observational constraints are not to be challenged, and remarkably these could be realised neatly by choosing appropriate scalar potential and/or coupling function, as in the chameleon \citep{kw, ms, lb2007, hs2007} and environment-dependent dilaton \citep{bcds2010} models.

An interesting property of these models is that either the mass of the scalar field or its coupling strength is sensitively dependent on the environments, in such a way that the fifth force is
suppressed where the observations and measurements are made. Taking the chameleon model as an example: in high density regions the scalar field becomes very massive so that
scalar field quanta could not propagate far and the fifth force gets suppressed. Exactly how massive the scalar field will be is determined by the local matter density as well as the steepness of the scalar potential and its derivatives, and the model could be so designed that the fifth force is suppressed in solar system but not on galactic and larger scales \citep{hs2007}, leaving the possibility that the structure formation process could be significantly affected. 

As one wants the scalar field mass to fluctuate strongly across the space, high degree of nonlinearity is inevitable. In chameleon model this is often reflected in the fact that the scalar field potential is very nonlinear, and in such circumstances linear treatment obviously fails. $N$-body simulations are then the natural method to be used to study structure formation in these models, and in this paper we shall apply this very technique to study the void properties of them.

Voids \citep{wp2009} are an important ingredient of the standard picture of hierachical structure formation. Because the initial matter distribution in the Universe is inhomogeneous, as time passes by the over-dense regions will pull more matter towards them and under-dense regions get emptied to form voids. For the $\Lambda$CDM cosmology, void phenomenon \citep{p2001} are well studied \citep{tc2009}. In coupled scalar field models, the fifth force, if unsuppressed, will boost the clustering of matter and therefore help to evacuate under-dense regions more quickly. A higher degree of emptiness in voids than what the concordance $\Lambda$CDM model predicts is thus an indicator of a possible fifth force \citep{knp2010}. The voids are of even greater importance to chameleon models because, by naive expectation, the fifth force in these models is suppressed in high density regions and shall not affect the galaxies clusters much, while in voids they are stronger and their effects more significant. It is therefore interesting to see what distinct features the voids have in these models compared with those in $\Lambda$CDM. As a preliminary work, we shall only consider dark matter voids \citep{csdgy} here, and leave the more technical work involving baryons to the future.

In this work we will investigate two coupled scalar field models, one in which the scalar field is a chameleon and the other in which the scalar field is not. In practice, there is no sharp distinction between them. For the scalar potential described by equation (\ref{eq:pot_nonchameleon}) below, for example, the scalar field has no chameleon features if parameter $\alpha\sim\mathcal{O}(0.1-1)$, but will become a chameleon while $\alpha\ll1$. Therefore, although for the non-chameleon models the fifth force is unsuppressed and thus could be well approximated as being proportional to gravity, as is commonly assumed by most $N$-body simulation works for coupled scalar field models to date (to name a few, \citet{mqmab2004, kk2006, fr2007, hj2009, knp2009, bprs2010}), for safety we shall solve the scalar field value as a function of spatial position explicitly and then differentiate to compute the fifth force \citep{lz2009, lz2010, lb2010a, lb2010b} (this %technique has recently been applied to the scalar tensor \citep{lmb2010} and the varying fundamental constant \citep{lbm2010} theories as well). 
technique has recently been applied to varying fundamental constant \citep{lbm2010} theories as well). 

The arrangement of this paper is as follows: in \S~\ref{sect:eqns} we give a brief summary of the major ingredients of the coupled scalar field model. In \S~\ref{sect:models} we explicitly write down the models we are simulating and present a detailed description of our void-finding algorithm. \S~\ref{sect:results} contains all our numerical results and finally \S~\ref{sect:con} is devoted to a summary and conclusions.

\section{The Basic Equations}

\label{sect:eqns}

All the equations relevant for the simulations here are derived
and discussed in \citep{lz2009, lz2010, lb2010a} but for the present
work to be self-contained we shall still list the minimum set of
them which is necessary for us to understand the underlying physics.

Instead of writing down the field equations directly as in some
previous work, we start from a Lagrangian density
\begin{eqnarray}\label{eq:lagrangian}
L =
{1\over{2}}\left[{R\over\kappa}-\nabla^{a}\varphi\nabla_{a}\varphi\right]
+V(\varphi) - C(\varphi)L_{\rm{DM}} +
L_{\rm{S}}
\end{eqnarray}
in which $R$ is the Ricci scalar, $\kappa=8\pi G$ with $G$ being the
gravitational constant, $L_{\rm{DM}}$ and
$L_{\rm{S}}$ are respectively the Lagrangian
densities for dark matter and standard model fields. $\varphi$ is
the scalar field and $V(\varphi)$ its potential; the coupling
function $C(\varphi)$ characterises the coupling between $\varphi$
and dark matter. Given the functional forms for $V(\varphi)$ and $C(\varphi)$
a coupled scalar field model is then fully specified.

Varying the total action with respect to the metric $g_{ab}$, we
obtain the following expression for the total energy momentum
tensor in this model:
\begin{eqnarray}\label{eq:emt}
T_{ab} = \nabla_a\varphi\nabla_b\varphi -
g_{ab}\left[{1\over2}\nabla^{c}\nabla_{c}\varphi-V(\varphi)\right]+ C(\varphi)T^{\rm{DM}}_{ab} + T^{\rm{S}}_{ab}
\end{eqnarray}
where $T^{\rm{DM}}_{ab}$ and $T^{\rm{S}}_{ab}$ are the
energy momentum tensors for (uncoupled) dark matter and standard
model fields. The existence of the scalar field and its coupling
change the form of the energy momentum tensor, and thus could modify the
cosmology from background expansion to structure formation.

Meanwhile, the coupling to scalar field produces a direct
interaction (a.k.a.~the fifth force) between dark matter
particles, due to the exchange of scalar quanta. This is best
illustrated by the geodesic equation for dark matter particles
\begin{eqnarray}\label{eq:geodesic}
{{d^{2}\bf{r}}\over{dt^2}} = -\vec{\nabla}\Phi -
{{C_\varphi(\varphi)}\over{C(\varphi)}}\vec{\nabla}\varphi
\end{eqnarray}
where $\bf{r}$ is the position vector, $t$ the (physical)
time, $\Phi$ the Newtonian potential and $\bf{\nabla}$ is the
spatial derivative. $C_\varphi=dC/d\varphi$. The second term in
the right hand side is the fifth force and only exists for coupled
matter species (dark matter in our model). The fifth force also
changes the clustering properties of the dark matter. Note that on
very large scales $\varphi$ could be considered as homogeneous and the fifth force
vanishes.

It has become obvious that in order to implement the above two
equations numerically we need to solve both the time evolution and
the spatial distribution of $\varphi$, and this could be done
using the scalar field equation of motion
\begin{eqnarray}
\nabla^{a}\nabla_a\varphi + {{dV(\varphi)}\over{d\varphi}} +
\rho_{\rm{DM}}{\frac{dC(\varphi)}{d\varphi}} = 0
\end{eqnarray}
or equivalently
\begin{eqnarray}
\nabla^{a}\nabla_a\varphi + {{dV_{eff}(\varphi)}\over{d\varphi}} =
0
\end{eqnarray}
where we have defined
\begin{eqnarray}
V_{eff}(\varphi) = V(\varphi) + \rho_{\rm{DM}}C(\varphi).
\end{eqnarray}
The background evolution of $\varphi$ can be solved easily once we
know the current $\rho_{\rm{DM}}$, because
$\rho_{\rm{DM}}\propto a^{-3}$. We can then divide $\varphi$
into two parts, $\varphi=\bar{\varphi}+\delta\varphi$, where
$\bar{\varphi}$ is the background value and $\delta\varphi$ is its
(not necessarily small nor linear) perturbation, and subtract the
background part of the scalar field equation of motion from the full equation
to obtain the equation of motion for $\delta\varphi$. In the
quasi-static limit in which we can neglect time derivatives of
$\delta\varphi$ as compared with its spatial derivatives (which
turns out to be a good approximation for our simulations, where
the simulation box is much smaller than the observable Universe),
we get
\begin{eqnarray}\label{eq:scalar_eom}
\vec{\nabla}^{2}\varphi =
{{dC(\varphi)}\over{d\varphi}}\rho_{\rm{DM}} -
{{dC(\bar{\varphi})}\over{d\bar{\varphi}}}\bar{\rho}_{\rm{DM}} +
{{dV(\varphi)}\over{d\varphi}} -
{{dV(\bar{\varphi})}\over{d\bar{\varphi}}}
\end{eqnarray}
where $\bar{\rho}_{\rm{DM}}$ is the background dark matter
density.

With the $\rho_{\rm{DM}}$ made ready on some grid, we could then solve
$\delta\varphi$ on that grid using a nonlinear Gauss-Seidel
relaxation method (in our simulations we have modified {\tt MLAPM},
 a public-available $N$-body code using a
self-adaptive refined grid so that high resolutions could be
achieved in high density regions). Because $\bar{\varphi}$ is also
known, we then get the full solution of
$\varphi=\bar{\varphi}+\delta\varphi$. This then completes the
computation of the source term for the Poission equation
\begin{eqnarray}\label{eq:poisson}
\vec{\nabla}^{2}\Phi =
{{\kappa}\over{2}}\left[C(\varphi)\rho_{\rm{DM}} -
C(\bar{\varphi})\bar{\rho}_{\rm{DM}} + \delta\rho_{\rm{B}}
- 2\delta V(\varphi)\right]
\end{eqnarray}
where
$\delta\rho_{\rm{B}}\equiv\rho_{\rm{B}}-\bar{\rho}_{\rm{B}}$
and $\delta V(\varphi)\equiv V(\varphi)-V(\bar{\varphi})$ are
respectively the density perturbations of baryons and scalar field
(note that we have neglected perturbations in the kinetic energy
of the scalar field because it is always very small for our
model).

We could then solve equation (\ref{eq:poisson}) using a linear
Gauss-Seidel relaxation method on the same grid to obtain $\Phi$.
With both $\Phi$ and $\varphi$ at hand, equation (\ref{eq:geodesic})
could be used to compute the forces on the dark matter particles,
and once we have the forces, we could do all the standard $N$-body
operations such as momentum-kick, position-drift, time-stepping
and so on.

Equations (\ref{eq:emt} - \ref{eq:poisson}) are all what we need to
complete an $N$-body simulation for coupled scalar field cosmology, 
and from then we could identify where the effects of
the scalar-coupling come in:
\begin{enumerate}
    \item The influence of the modified cosmic background expansion rate mainly
    comes through the particle movements and time-stepping, {\it i.e.},
    equation (\ref{eq:geodesic}). This is because in the simulations we
    shall use scale factor $a$ as the time variable and
    $d/dt=\dot{a}d/da$.
    \item The varying mass effect could be seen directly from
    equation (\ref{eq:poisson}), which shows that the contribution of
    the dark matter density $\rho_{\rm{DM}}$ to the source of the Poisson
    equation is multiplied by a factor $C(\varphi)$ which differs from
    1 in general. In our model the mass of
    dark matter particles is not really varying, but the net effect is
    just that.
    \item The fifth force appears explicitly in the right hand
    side of the geodesic equation (\ref{eq:geodesic}), but only for coupled
    matter species (dark matter in our model).
    \item The velocity-dependent "frictional force" is a bit
    subtler. It hides behind the fact that equation (\ref{eq:geodesic}),
    equation (\ref{eq:scalar_eom}) are actually written in {\it different}
    gauges: equation (\ref{eq:geodesic}) is the force for a dark matter
    particle and is given in that particle's rest frame, while
    equation (\ref{eq:scalar_eom}) is written in the fundamental
    observer's frame. As a result, to use the $\delta\varphi$ solved
    from equation (\ref{eq:scalar_eom}) in equation (\ref{eq:geodesic})
    we need to perform a frame transform
    $\vec{\nabla}\delta\varphi\rightarrow\vec{\nabla}\delta\varphi+a\dot{\bar{\varphi}}\dot{\bf{x}}$
    in which $\dot{\bf{x}}$ the comoving velocity of the said
    particle relative to the fundamental observer. This force is
    thus expressed as
    $-{{C_\varphi}\over{C}}a\dot{\varphi}\dot{\bf{x}}$, and
    obviously the faster a particle travels the stronger
    frictional force it feels.
\end{enumerate}

In our numerical simulation, we have included all these effects consistently \citep{lb2010a}. In particular, we have computed
the fifth force explicitly, rather than simply assuming that it is always proportional to gravity: as shown
in \citet{lz2009, lz2010}, such assumption could be fairly poor for certain models where the scalar
field configuration is very inhomogeneous, although it is good enough for other models \citep{lb2010a}.

\section{Simulations and Void Finding Algorithm}

\label{sect:models}

\subsection{The Models Studied}

In this work we consider the voids in the two different coupled scalar field models studied respectively by
\citet{lb2010a} and \citet{lz2009, lz2010, zmlhf}. Both models have an exponential coupling between dark
matter and the scalar field $\varphi$, 
\begin{eqnarray}\label{eq:coupling}
C(\varphi) = \exp(\gamma\sqrt{\kappa}\varphi)
\end{eqnarray}
and run-away potentials for $\varphi$:
\begin{eqnarray}\label{eq:pot_nonchameleon}
V(\varphi) = {{\Lambda}\over{\left(\sqrt{\kappa}\varphi\right)^\alpha}}
\end{eqnarray}
for the model of \citet{lb2010a}, and
\begin{eqnarray}\label{eq:pot_chameleon}
V(\varphi) = {{\Lambda}\over{\left[1-\exp\left(-\sqrt{\kappa}\varphi\right)\right]^\alpha}}
\end{eqnarray}
for the model of \citet{lz2009, lz2010}. In the above $\Lambda$ is a parameter of mass dimension 4 and 
is of order the present density for dark energy ($\varphi$ plays the role of dark energy in the models); 
its precise value is determined by the numerical code for the consistency
in the background cosmology \citep{lb2010a}. $\gamma$ and $\alpha$ are dimensionless 
model parameters controlling respectively the strength of the coupling and the steepness of the potentials. 

For the potential equation (\ref{eq:pot_nonchameleon}) we choose $\alpha=0.1$ and $\gamma<0$ so that the total effective potential 
$V_{eff}(\varphi)$ is of runaway type. The scalar field then rolls quickly at early times ($\varphi_{\rm today}-\varphi_{\rm early}
\sim \mathcal{O}(M_{\rm Pl})$ with $M_{\rm Pl}$ the Planck mass), until slowing down to the slow-roll regime, when it
behaves like a normal quintessence field. For the potential in equation (\ref{eq:pot_chameleon}) we choose $\alpha\ll1$ and $\gamma>0$,
which ensures that $V_{eff}$ has a global minimum close to $\varphi=0$ and $d^2V_{eff}(\varphi)/d\varphi^2
\equiv m^2_{\varphi}$ at this minimum is very large in high density regions; then $\varphi$ is trapped close to 0 
all through the cosmic history. These two cases are two extremes of the coupled scalar field field: in the former
$\varphi$ clusters very weakly just as in normal quintessence, and the fifth force is, to a good approximation, 
always proportional to gravity; the scalar field coupling also drastically modifies the background cosmology and
structure formation at early times ($z\sim \mathcal{O}(10^2)$). In the latter case $\varphi$ is very inhomogeneous and the
fifth force is greatly suppressed in high density regions where $\varphi$ acquires heavy mass, $m^2_{\varphi}\gg 
H^2$ ($H$ is the Hubble expansion rate), and thus the fifth force cannot propagate far. The suppression of the
fifth force is even severer at early times, meaning that the structure formation is only influenced at late times
($z$ less than a few); also, because $\varphi$ is trapped close to 0 all the time, the background cosmology is
forced to be indistinguishable from $\Lambda$CDM. In Table 1 we summarise the details for all the models we 
study here \footnote{For clearness we shall refer to the model with equation (\ref{eq:pot_chameleon}) as chameleon model, because
here the scalar field mass $m_{\varphi}$ depends sensitively on its environment and fluctuates strongly from point 
to point, and correspondingly the model with equation (\ref{eq:pot_nonchameleon}) as the non-chameleon model because here $m_{\varphi}$
is largely ignorant of the environment. The formal definition of a chameleon model \citep{kw, ms}  is not relevant for our discussion in this work.}.
\begin{table}
\caption{A summary of the details of the models studied here. For all the runs:
$\Omega_m=0.257$, $n=0.963$, $\sigma_8=0.769$, $H_0=71.9$ km/s/Mpc; size of simulation 
box is 64$h^{-1}$~Mpc and particle number is $256^3$ so that mass resolution is $1.04\times10^9h^{-1}~M_{\sun}$; 
domain grid has $128$ cells in each side, and the finest refinement has $16384$ cells in each side, leading to
a force resolution of $\sim12h^{-1}$~kpc. L, N, C stand for $\Lambda$CDM, non-chameleon and chameleon
respectively.}
\begin{tabular}{@{}lccccc}
\hline\hline
model & $V(\varphi)$ & $C(\varphi)$ & $\gamma$ & $\alpha$ & $\kappa\Lambda/3H_0^2$\\
\hline
L	& constant & $1$ & $0.0$ & $0.0$ & $0.743$ \\
N1	& eqn (\ref{eq:pot_nonchameleon}) & eqn (\ref{eq:coupling}) & $-0.05$ & $0.1$ & $0.717$ \\
N2	& eqn (\ref{eq:pot_nonchameleon}) & eqn (\ref{eq:coupling}) & $-0.10$ & $0.1$ & $0.781$ \\
N3	& eqn (\ref{eq:pot_nonchameleon}) & eqn (\ref{eq:coupling}) & $-0.15$ & $0.1$ & $0.838$ \\
N4	& eqn (\ref{eq:pot_nonchameleon}) & eqn (\ref{eq:coupling}) & $-0.20$ & $0.1$ & $0.891$ \\
C1	& eqn (\ref{eq:pot_chameleon}) & eqn (\ref{eq:coupling}) & $0.5$ & $1.0\times10^{-6}$ & $0.743$ \\
C2	& eqn (\ref{eq:pot_chameleon}) & eqn (\ref{eq:coupling}) & $0.5$ & $1.0\times10^{-5}$ & $0.743$ \\
C3	& eqn (\ref{eq:pot_chameleon}) & eqn (\ref{eq:coupling}) & $1.0$ & $1.0\times10^{-6}$ & $0.743$ \\
C4	& eqn (\ref{eq:pot_chameleon}) & eqn (\ref{eq:coupling}) & $1.0$ & $1.0\times10^{-5}$ & $0.743$ \\
\hline
\end{tabular}
\end{table}

\subsection{Void Finding Algorithm}

Following \citet{csdgy}, our void finding algorithm consists of two steps: the
identification of spherical proto-voids and mergers of proto-voids to form voids of arbitrary shape (all through this paper proto-voids and voids are different things and are not to be confused). 

The proto-voids are spherical regions in which the average of the density contrast $\delta=
\rho/\bar{\rho}-1$ is below some predefined threshold $\delta_v$. As shown by \citet{csdgy}, 
voids very clearly correspond to the troughs of the initial density field, justifying the assumption that
voids grow gravitationally from the initial negative overdensities. Assuming a spherical evolution model
for the voids \citep{gg, ddglp}, the growth of the voids can be studied 
analytically, and it is found that at the time of shell-crossing the overdensity inside the spherical proto-void
reaches $-0.8$. Although this is the result for Einstein-de~Sitter cosmology \citep{csdgy}, we 
shall adopt it as a guidance and set $\delta_v=-0.8$ in the coupled scalar field models as well.

Our void finding algorithm is similar to that of \citet{csdgy}, but it differs from the latter in various
details, particularly the treatment of the mergers of proto-voids. To be clear and self-contained, here we
briefly describe our algorithm in separate steps:

(1) A regular $128\times128\times128$ mesh is set up and the particle densities on this mesh are computed
using the Triangular-shaped Cloud ({\tt TSC}) scheme. This scheme ensures that the density interpolation
is smoother than the usually used Cloud-in-Cloud scheme.

(2) The local minima in the density field are located, and these are considered as the centres of the proto-voids 
\citep{wk1993}. Top-hat spherical windows with large enough radii so that the smoothed density 
contrasts inside are greater than $\delta_v$ are then placed at these minima, and the radii are gradually decreased
until the density contrast drops below $\delta_v$. These minima and radii are then taken as the centres and sizes of the
proto-voids respectively. One can also do this for all grid points on our mesh ({\it i.e.,} set up a top-had window
on each grid point, decrease the radii of the windows until the overdensity at a grid point falls below $\delta_v$) as
in \citet{knp2010}, but this is more time-consuming and we have checked that the two methods lead to 
compatible results.

(3) The above identified proto-voids are merged as appropriate to form the final voids of arbitrary shapes. It
is well known that voids occupy the majority of the space, with islands of matter (dark matter halos and galaxies) interconnected by
the narrow filaments which go through them. As a result, the merging criteria must be chosen carefully: for example,
the dumbbell-shaped configurations are better to be avoided \citep{csdgy} to prevent the proto-voids
from all being merged to form a single void as big as the simulation box. To this end let's adopt a variant of the merging
criterion proposed by \citet{csdgy}, which consists of the following steps:
\begin{enumerate}
\item Consider all the proto-voids which have not been assembled into any final voids. Find the biggest one,
which is the primary progenitor of a merged void-to-be, and all the smaller ones which intersect it.
\item If a smaller proto-void fully lies within the biggest one, then it is removed from the list.
\item {\it All} the smaller proto-voids whose centres lie inside the biggest one are merged to the latter.
\item {\it All} the smaller proto-voids whose centres lie outside the biggest one while still have significant intersection with
the latter are merged to the latter. There are certain freedoms as for what is to be considered as significant, and here we will
adopt the proposal of \citet{csdgy}: we divide the line segment connecting the centres of the smaller and biggest
proto-voids into three sections, a section $a$ which lies in both spheres and two sections $b, c$ lying only in one of the two
spheres respectively. The intersection is only considered to be significant if $|a| \gid \max(|b|, |c|)$. 
\item {\it All} the smaller proto-voids which do not satisfy $|a| \gid \max(|b|, |c|)$ are {\it not} merged into the biggest one. 
In this case, the portion of volume shared by the smaller and the biggest proto-voids is assigned to the latter, and the volume
of the smaller proto-void is decreased correspondingly. The smaller proto-voids are however not removed from the list but
will be the building blocks for the voids-to-be considered later (as are the proto-voids which have no intersections with the 
currently biggest one).
\item The total volume of the final merged void is that of the union of the biggest proto-void and {\it all} those smaller ones 
merged to (eaten by) it\footnote{The volume is computed as follows: we distribute a big number of points
evenly in the whole volume so that the number density $n$ is known, and then count the number $N$ of points which lie inside the 
union. The volume of the union is then simply $N/n$. With our choices of the number density, we find that the numerical results 
agree with analytical predictions (where the latter are available) typically better than $99.9999\%$. We are also careful to avoid 
assigning a same portion of space to more than one merged voids: for example, when computing the volume of a void-to-be, we always exclude
the parts which lie in any of the previously identified voids.}. By our construction the shape of the final void could be arbitrary, but as 
\citet{csdgy} we define an effective radius $r_{eff}$ such that ${{4\pi}\over3}r^2_{eff}$ equals the volume of it. The
centre of the merged void, on the other hand, is taken to be the volume average of the centres of all merging blocks. 
\item The biggest proto-void and all the smaller ones eaten by it (but {\it not} the ones dealt with in step (v) above) are considered 
to be already assembled, and are excluded from the future runs of the merging process. Steps (i) - (vii) are then repeated
until all the proto-voids are assembled.
\end{enumerate}
Note that using this algorithm we could naturally avoid the the dumbbell-shaped configurations for which two big proto-voids are
connected by thin tunnels and are finally merged, because the two big proto-voids have no intersection by definition, and are thus 
treated separately. The multiple merges, in contrast with the algorithm of \citet{csdgy} which stops looking for further
overlaps one it has found one, could produce bigger big voids, for which the biggest proto-voids typically eat tens of smaller ones, 
thereby expanding significantly.

In this work we only consider voids whose effective radius $r_{eff}\geq2.0h^{-1}$~Mpc, which is 4 times the size of the cells in 
the void-finder grid \citep{csdgy}. As the majority of the empty space is populated by the very small voids, this will inevitably leave part of
the void space undetected, as a price of maintaining accuracy.

\section{Numerical Results}

\label{sect:results}

\begin{figure*}
\includegraphics[width=180mm]{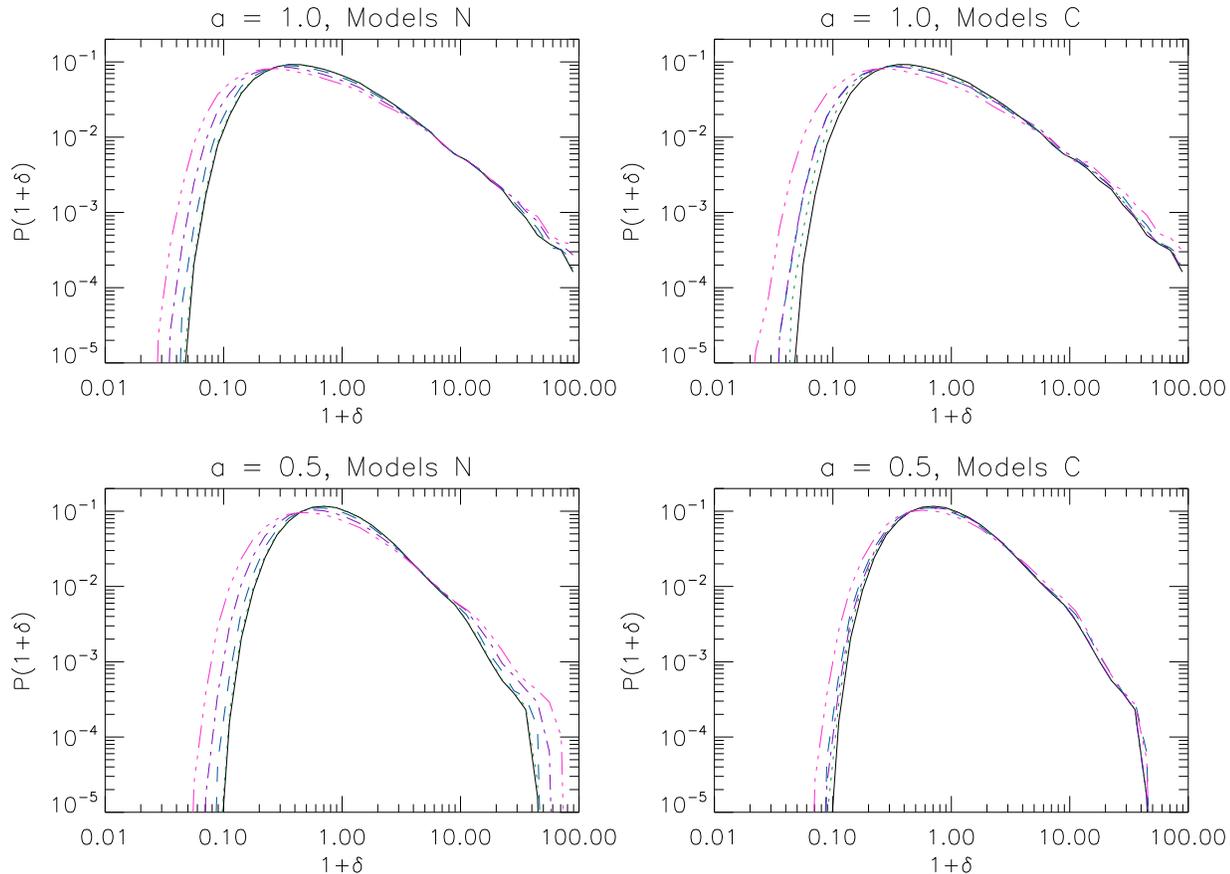}
\caption{The probability distribution of the density contrast $\delta$ (equivalently $1+\delta$). {\it Upper Left Panel}: Results at redshift 0 for
the non-chameleon models N1 (green dotted curve), N2 (blue dashed curve), N3 (purple dot-dashed curve) and N4 
(pink dot-dot-dot-dashed curve) in contrast to the $\Lambda$CDM (L) result (black solid curve). {\it Upper Right Panel}: The same
but the green dotted, blue dashed, purple dot-dashed and pink dot-dot-dot-dashed curves now represent respectively the models
C1, C2, C3 and C4. {\it Lower Left Panel}: The same as the upper left panel but for redshift 1. {\it Lower Right Panel}: The same as
the upper right panel, but for redshift 1.}
\label{densprob}
\end{figure*}

\begin{figure*}
\includegraphics[width=180mm]{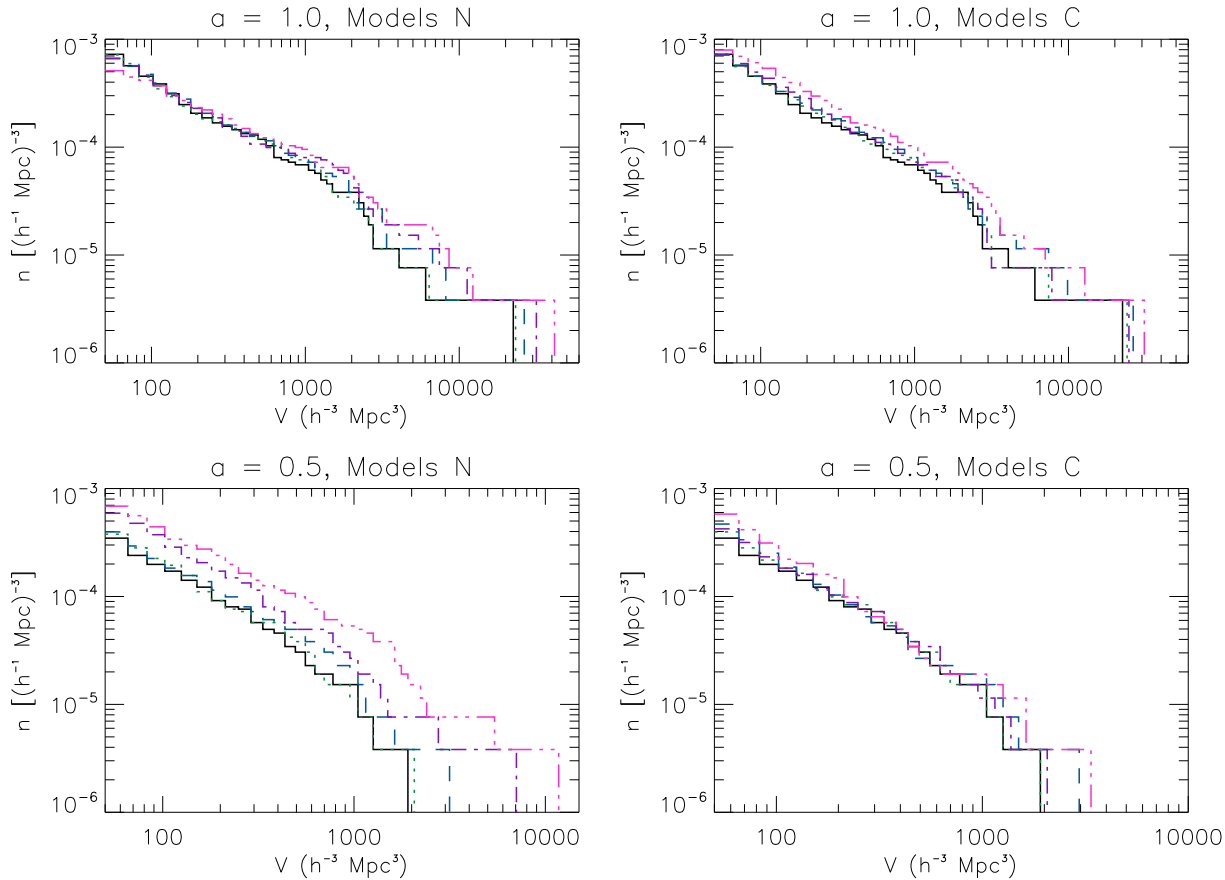}
\caption{The void volume functions for the simulated models. {\it Upper Left Panel}: Results at redshift 0 for the non-chameleon models N1 
(green dotted curve), N2 (blue dashed curve), N3 (purple dot-dashed curve) and N4 (pink dot-dot-dot-dashed curve) in comparison to model 
L (black solid curve). {\it Upper Right Panel}: The same, but the green dotted, blue dashed, purple dot-dashed and pink dot-dot-dot-dashed
curves now stand for model C1, C2, C3 and C4 respectively. {\it Lower Left Panel}: The same as the upper left panel but for redshift 1. {\it Lower
Right Panel}: The same as the upper right panel, but for redshift 1. In all the plots the horizontal axis is the void volume $V$, in unit of 
$\left(h^{-1}{\rm Mpc}\right)^3$, and the vertical axis is the number density of voids which are larger than $V$, in unit of 
$\left(h^{-1}{\rm Mpc}\right)^{-3}$.}
\label{vvf}
\end{figure*}

\begin{figure*}
\includegraphics[width=180mm]{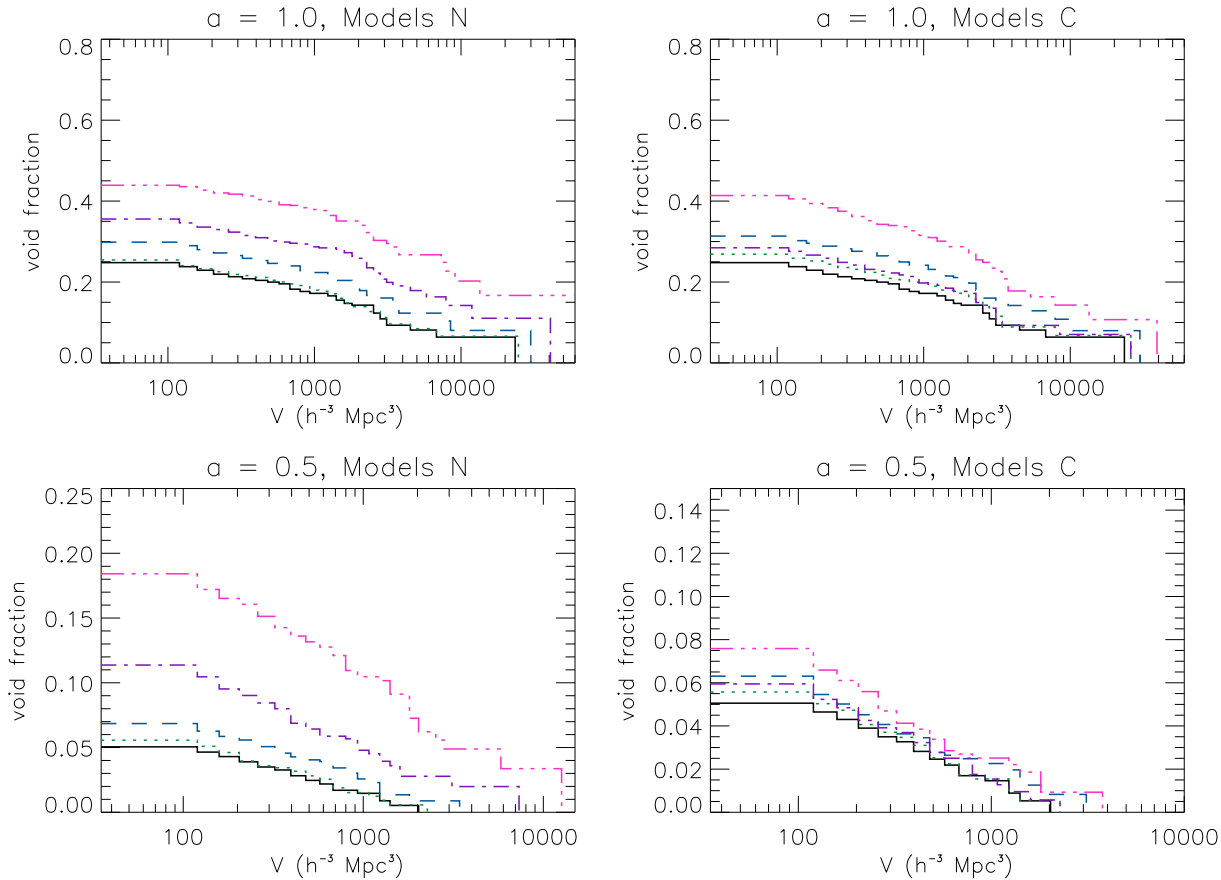}
\caption{The fraction of the total space that is occupied by voids larger than $V$ as a function of $V$. {\it Upper Left Panel}: Results at redshift 0 for the non-chameleon models N1 
(green dotted curve), N2 (blue dashed curve), N3 (purple dot-dashed curve) and N4 (pink dot-dot-dot-dashed curve) in comparison to model 
L (black solid curve). {\it Upper Right Panel}: The same, but the green dotted, blue dashed, purple dot-dashed and pink dot-dot-dot-dashed
curves now stand for model C1, C2, C3 and C4 respectively. {\it Lower Left Panel}: The same as the upper left panel but for redshift 1. {\it Lower
Right Panel}: The same as the upper right panel, but for redshift 1. In all the plots the horizontal axis is the void volume $V$, in unit of 
$\left(h^{-1}{\rm Mpc}\right)^3$, and the vertical axis is the fraction of space occupied by voids larger than $V$.}
\label{ff}
\end{figure*}

\begin{figure*}
\includegraphics[width=180mm]{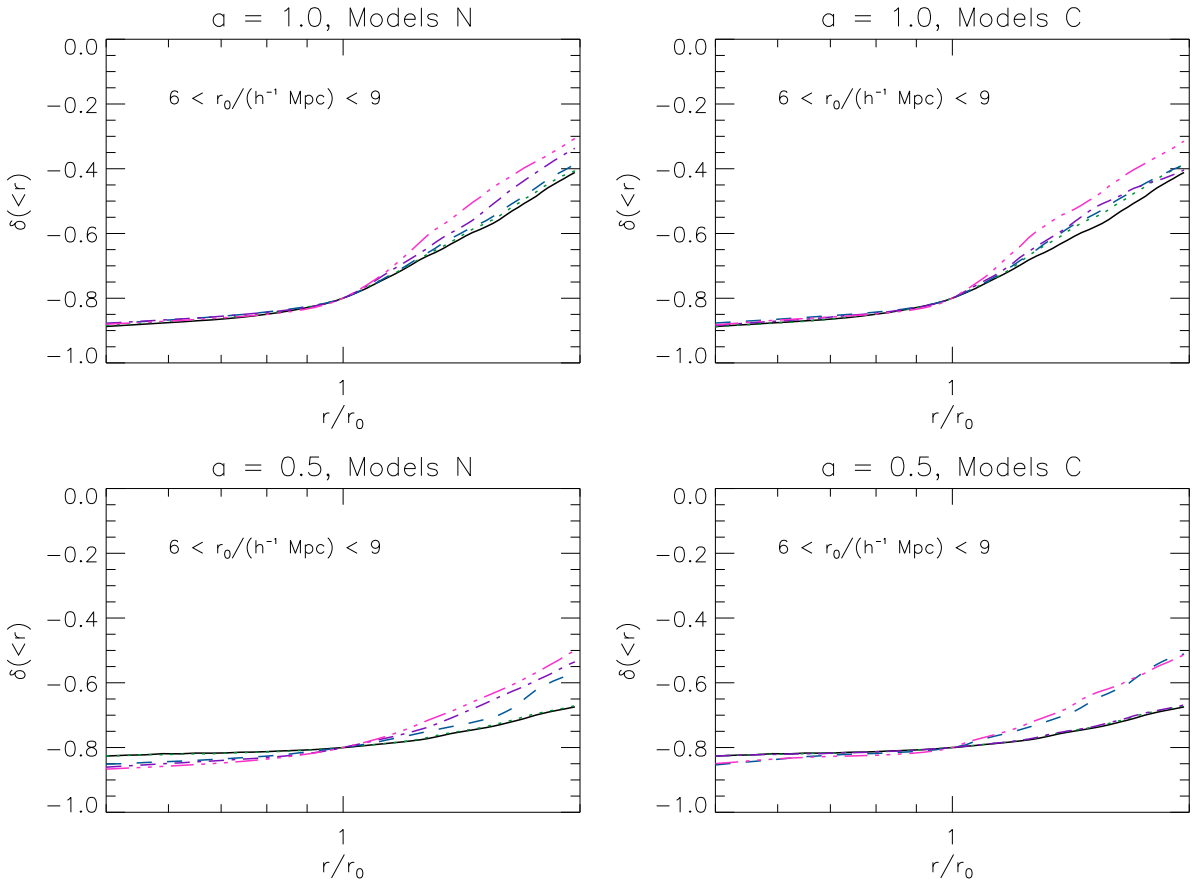}
\caption{The density profiles of the proto-voids whose radii fall in the range $6\leq r_0/\left(h^{-1}~{\rm Mpc}\right)\leq9$. {\it Upper Left Panel}: 
Results at redshift 0 for the non-chameleon models N1 
(green dotted curve), N2 (blue dashed curve), N3 (purple dot-dashed curve) and N4 (pink dot-dot-dot-dashed curve) in comparison to model 
L (black solid curve). {\it Upper Right Panel}: The same, but the green dotted, blue dashed, purple dot-dashed and pink dot-dot-dot-dashed
curves now stand for model C1, C2, C3 and C4 respectively. {\it Lower Left Panel}: The same as the upper left panel but for redshift 1. {\it Lower
Right Panel}: The same as the upper right panel, but for redshift 1. In all panels the horizontal axis is the distance from the proto-void centre, $r$,
 in unit of the radius of the proto-voids, $r_0$; the vertical axis is the dimensionless density contrast.}
\label{vprofile_l}
\end{figure*}

\begin{figure*}
\includegraphics[width=180mm]{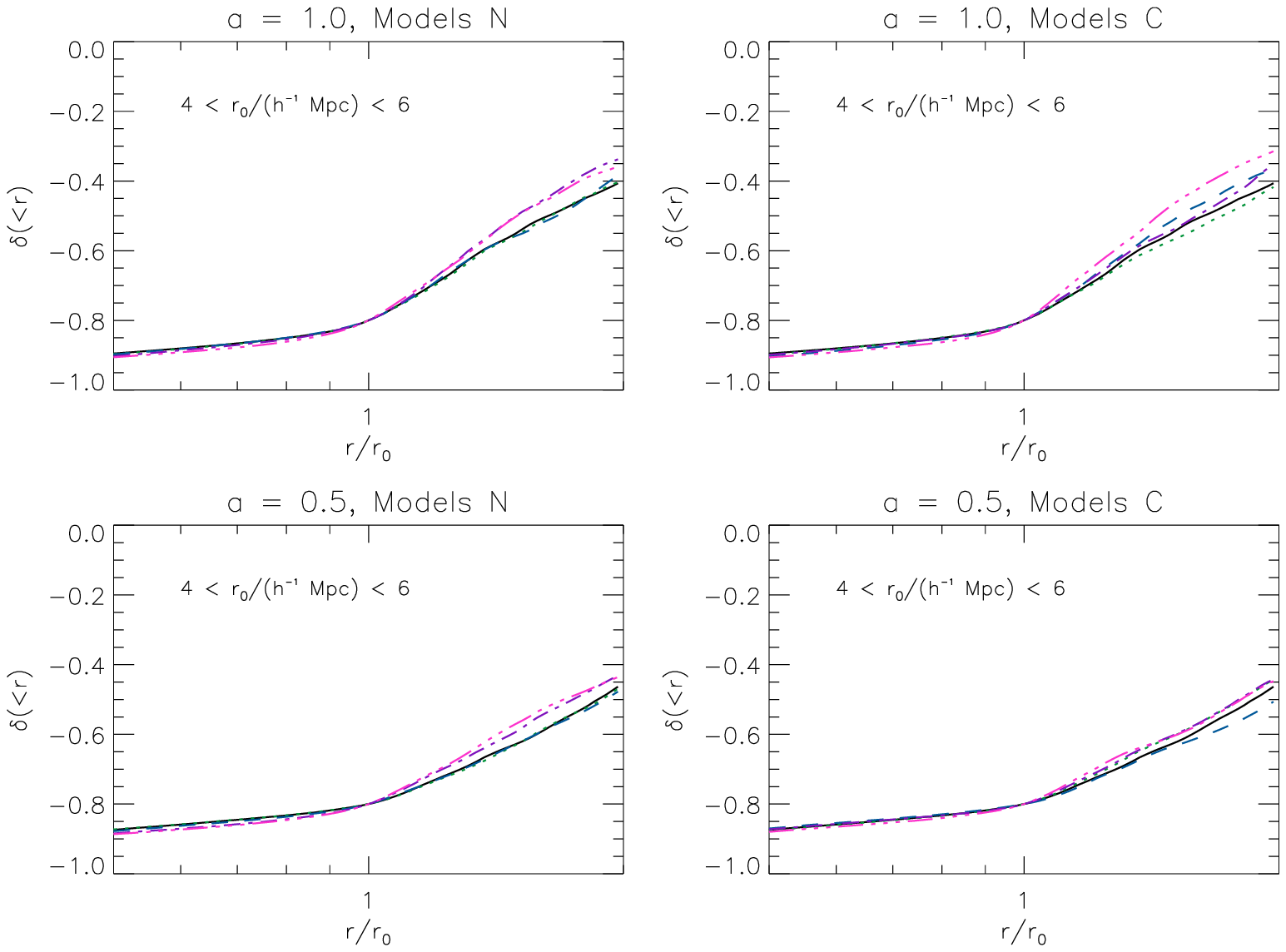}
\caption{The same as Fig.~\ref{vprofile_l}, but for proto-voids whose radii fall into the interval $4\leq r_0/\left(h^{-1}~{\rm Mpc}\right)\leq6$.}
\label{vprofile_s}
\end{figure*}

Having introduced the models and void finding algorithm in the above section, we now present and discuss the numerical results.

\subsection{Probability Distribution of Density Contrast}

As is shown in \citet{lb2010a, lb2010b} and \citet{lz2009, lz2010}, the existence of the scalar field and its coupling to dark matter efficiently 
enhance the clustering of matter, and we thus expect that an associated effect is the fast evacuation of the low density regions, 
resulting in a bigger portion of the space being inside voids (than in $\Lambda$CDM model). 

To give a first check of this expectation, we consider the probability distribution of the density contrast,
which shows how much of the total space is in overdense/underdense regions. To obtain this, we place top-hat windows with
radius $2.0h^{-1}$~Mpc at each of our $128^3$ regular grid points and compute the smoothed density inside them. We then count
the number of grid points at which the density contrast $\delta$ falls within $[\delta_0,\delta_0+d\delta]$, and divide this by $128^3$ 
to obtain the probability that $\delta\in[\delta_0,\delta_0+d\delta]$.

The results for our 9 models are summarised in Fig.~\ref{densprob} (see the figure caption for a detailed description), from which we 
could see clearly the trend that the coupling scalar field models (both N and C) predict more higher and lower density regions than 
$\Lambda$CDM (L). However, there are some notable differences between them, reflecting the fact that the N and C models are 
qualitatively different. 

For the C models, as mentioned above, the scalar field $\varphi$ is trapped at the minimum of $V_{eff}$, $\varphi_\ast$, which is close to 0 all 
through the cosmic history (as can be easily checked using the equation below). Using this fact it could be straightforward to find that
\[
\sqrt{\kappa}\varphi_\ast \approx \frac{\alpha\Lambda}{\gamma\rho_{\rm DM}},
\]
and thus \citep{lz2009}
\begin{equation}\label{eq:scalar_mass}
m^2_{\varphi} = \frac{\partial^2V_{eff}}{\partial\varphi^2}(\varphi_\ast) \approx \frac{\left(\gamma\kappa\rho_{\rm DM}\right)^2}{\alpha\kappa\Lambda}.
\end{equation}
Although in reality the fifth force is quite complicated because the force between two particles depends on the matter distribution in between them, 
the quantity $m_{\varphi}$ nevertheless could be utilised to qualitatively understand the underlying physics. Basically, the greater $m_{\varphi}$ is,
the shorter the distance the fifth force could propagate, which means that a particle will feel the fifth forces exerted by less particles: in short, a heavy
mass $m_{\varphi}$ could {\it suppress} the fifth force. 

From equation (\ref{eq:scalar_mass}), we see that $m_{\varphi}$ increases as $\gamma, \rho_{\rm DM}$ increase or $\alpha$ decreases, and vice
versa. Thus we expect the fifth force to be more severely suppressed in models C1, C3 in which $\alpha$ is smaller, than in models C3, C4; within C1, C3
it is more suppressed for C3 for which $\gamma$ is bigger, and for all models it is more suppressed at higher redshift, where $\rho_{\rm DM}$ is bigger
overall \citep{lz2009, lz2010}. Furthermore, $\rho_{\rm DM}$ is generally lower in the void regions, where the fifth force is expected to be less suppressed.

In Fig.~\ref{densprob} we could see that at $a=0.5$ (redshift 1, lower right panel) the probability distributions of the larger density contrasts for the models C1-C4
are essentially the same as in model L, because fifth force is strongly suppressed there and then. On the other hand, for lower density contrasts in the low density regions, we do see
the deviations from model L increasing for decreasing $m_{\varphi}$, as expected. As time passes, $m_{\varphi}$ decreases overall and by $a=1.0$ (redshift 0,
upper right panel) a significant greater portion of space will be evacuated in the C models than in the L model, and meanwhile the enhancement in the
clustering of matter due to the (attractive) fifth force increases the chance the peaks in the initial density field develop into highly overdense regions ($\delta\gg1$).

For the N models, equation (\ref{eq:scalar_mass}) does not apply because the scalar field always rolls down the effective potential $V_{eff}$. $m^2_{\varphi}$
does not fluctuate strongly in space, and the fifth force is never severely suppressed \citep{lb2010a}. As a result, the fifth force starts to affect the structure
formation at a fairly early time. As is shown in Fig.~\ref{densprob}, by $a=0.5$ (redshift 1, lower left panel) there have been significantly more overdense
{\it and} underdense regions than L model, mainly because the fifth force helps transfer more matter from low density to high density regions\footnote{Note that the probability is higher in the N models than in the L model for both $1+\delta\gg1$ and $1+\delta\ll1$, as the total amount of matter is the same in both models}.

At $a=1.0$, however, although the N models still predict larger evacuated space than model L does (upper left panel), the volumes of space in the very overdense
regions ($\delta\gg1$) are less different between  the N and L models, which is because as more and more matter is transferred to the high density regions, there is less and less
remaining in the empty regions to be pumped: even though the aggregation of matter into high density regions starts earlier in the N models, it slows down eventually
as matter in the low density regions is used up, and there turns out to be time for the L model to catch up somehow. These results for our coupled scalar field models (mainly the N models) is similar to that of the ReBEL model \citep{ngp2005} as investigated by \citet{knp2010} and \citet{hj2009}.

\subsection{Volume of Voids}

Having seen above that the fifth force in both the N and C models helps evacuate the low density regions, we now have a look at how the void properties are
affected.

The first interesting quantity is the void volume function (similar to the halo mass function in the studies of the statistical properties of dark matter halos), which shows
the number density of voids larger than a given volume $V$. Fig.~\ref{vvf} displays our results for the N, C and L models.

Understandably, the more strongly the matter particles cluster, the more effectively the low density regions are evacuated and therefore the bigger
the sizes of the voids tend to be. In the N models, not only does the fifth force, which is unsuppressed, start to take effect earlier, but also the universe 
expands more slowly, leaving more time for particles to clump \citep{lb2010a, lb2010b}. Consequently, by $a=0.5$ (Fig.~\ref{vvf}, lower left panel), we see
large increases in both the void number density and void size in the N models as compared to the L model (for example, the number density could be twice as high). 

Again, as time goes on, low density regions are largely emptied and few particles remain there, then the growth of the voids will slow down. The earlier the evacuation
starts, the earlier it will be completed and the growth of voids stops. As a result, when voids in the N models stop growing, those in the L model are still in the process. 
Finally, at $a=1.0$, the void volume function of the L model more or less catches up those in the N models, as is evident in the upper left panel of Fig.~\ref{vvf}. We 
also note that at $a=1.0$ there are less small voids in the N4 model than in the L and other N models, which is likely because of the fact that small voids have been used up
to merge to form bigger ones.

For the C models, the suppress of the fifth force means that the clustering of matter and growth of voids are less affected by it \footnote{One might think that, from 
equation (\ref{eq:scalar_mass}), $m_{\varphi}$ could be quite light in low density regions and thus the fifth force should be essentially unsuppressed there. The point, 
however, is that equation (\ref{eq:scalar_mass}) is only used to understand the physics intuitively, and in reality the fifth force is proportional to $\vec{\nabla}\delta\varphi$ 
[cf.~equation (\ref{eq:geodesic})] with $\delta\varphi$ determined by the dynamical equation (\ref{eq:scalar_eom}). As a result, the value of $\delta\varphi$ in void regions
generally depends on that outside the regions in the way that the solution to a differential equation depends on its boundary conditions.}. This is easily seen in the $a=0.5$ case 
(Fig.~\ref{vvf}, lower right panel), which shows that the void volume functions for the C models do not deviate much from that for the L model (one might appreciate the
effect of the suppress in the fifth force by considering that the ratio between the magnitudes of fifth force and gravity is $2\gamma^2$ if the former is not suppressed, and
$\gamma\sim\mathcal{O}(0.1)$ for N models while $\gamma\sim\mathcal{O}(1)$ for C models). 

We could also have an examination of the void filling factor, defined as the fraction of total space that is filled by voids which are either bigger or smaller than $V$.
Because our algorithm leaves the very small voids undetected, we choose to show the former, and the results are given in Fig.~\ref{ff}. It turns out that this plot shows
more clearly the effects of the scalar coupling. As our first example, for the L model at $a=0.5$ (Fig.~\ref{ff}, lower left panel), we notice that only 5\% of the total space is filled 
by voids larger than $35h^{-3}$~Mpc$^3$, in contrast to more than 11\% and 18\% for the models N3 and N4 respectively. At $a=1.0$, as a result of void growth and 
mergers, the numbers for these three models are changed to 25\%, 35\% and 45\% respectively. In both cases, the scalar field coupling dramatically changes the 
total volume of void regions, and could potentially affect the properties of void galaxies (though the resolution of our simulations is limited and so we shall not touch this in this work).

For the C models (Fig.~\ref{ff}, right panels), we obtain qualitatively similar results, but the deviations from model L are obviously smaller due to the suppress of the
fifth force. For example, at $a=0.5$, the fraction of space occupied by voids larger than $35h^{-3}$~Mpc$^3$ is respectively $5\%$, $6\%$ and $7.5\%$
for the models L, C3 and C4, while at $a=1.0$ these numbers become 25\%, 28\% and 41\%.

\subsection{Void Density Profiles}

The next quantity we are interested in is the void density profile, which characterises how matter is distributed within the empty regions. Like the growth rate of 
voids, this is also interesting and bears information about the physics driving the structure formation. In \citet{knp2010}, for example, it is shown that the ReBEL model
could produce different density profiles inside the void: the profile is steeper than the $\Lambda$CDM prediction, and is more so for smaller voids. In this subsection we 
would like to see what happens for our coupled scalar field models.

As our voids are made from the spherical proto-voids and thus in principle could have arbitrary shape, it's difficult to give well-defined profiles for them. Instead,
because we are only interested the steepness at the boundaries of the empty regions, here we choose to compute the profiles for the {\it proto-voids}, which are simply obtained by varying
the radii of the top-hat smoothing windows located at the centres of the proto-voids and calculating the average densities inside them.

As \cite{knp2010}, we'll consider two groups of the (proto-)voids, with radius ranges of $6\leq r_0/\left(h^{-1}{\rm Mpc}\right)\leq 9$ and $4\leq r_0/\left(h^{-1}{\rm Mpc}\right)\leq 6$
respectively, where $r_0$ is the radius of the proto-void.

For the group of larger voids, with radii in the range $6\leq r_0/\left(h^{-1}{\rm Mpc}\right)\leq 9$, the results are shown in Fig.~\ref{vprofile_l}. As in the ReBEL model, 
in both our N and C models the scalar field coupling makes the density profile deeper in the outer region of the proto-voids, and this effect is more prominent in early times 
(redshift 1.0, lower panels), which is not unexpected, because the fifth force helps to make the structure formation process start (and low density regions get evacuated) earlier, 
a fact which has also been confirmed by the observation that at early times the proto-voids have notably lower density in the very inner regions in the coupled scalar field models than in
$\Lambda$CDM. Note that the chameleon effect does play a role here, by making the proto-void density profiles in models C1 and C3 (for which $\alpha=10^{-6}$ and fifth force
most severely suppressed) indistinguishable from that of the L model; at later times (redshift 0.0, Fig.~\ref{vprofile_l}, upper right panel), however, the fifth force becomes less suppressed and the deviations from the L model gradually develop.

Fig.~\ref{vprofile_s} displays the same results, but for the group of smaller (proto-)voids. Here we can see the similar trend as in Fig.~\ref{vprofile_l}, namely the scalar coupling produces
steeper increase in the void density profile around the void edge. The effect is not as strong as for the large halos, possibly because smaller voids form earlier and have been effectively 
evacuated even if the fifth force is not at play. 

These results suggest that larger voids at earlier times could better reveal the influences of a possible scalar field coupling than smaller ones. Also, the
strong-chameleon and non-chameleon models can be distinguished because for the former the void profile is essentially the same as the $\Lambda$CDM prediction at earlier times but starts to deviate later, while for the latter the deviation starts quite early.

Because of the limitation of resolution, we are unable to test the void profiles for very large and very small voids, which will be left for future work. Given the fact that large voids could be very different in size in different models, we expect that their density profiles could reveal more information about the physical models.

\section{Discussion and Conclusion}

\label{sect:con}

Scalar-field-mediated long-range fifth forces have attracted much attention among cosmologists in recent years, and in most versions 
they come from a direct coupling between the matter species (usually dark matter only) and a cosmological scalar field. If they really exist, 
they might dramatically change the picture of cosmic structure formation, and alleviate or even solve some problems in the concordance 
$\Lambda$CDM model. On the other hand, the range and magnitude of the fifth force are often model dependent: in one extreme, 
represented by our N models, the effective potential $V_{eff}(\varphi)$ is fairly flat so that the mass $m_\varphi$ is light and almost the same everywhere; the
fifth force then has a fixed ratio to gravity, which is equivalent to a rescaling of the gravitational constant. In the opposite extreme, like our C models, $V_{eff}$ and thus
the scalar field mass $m_\varphi$ depends sensitively to the matter density in the way that $m_{\varphi}$ could be very heavy in high density regions, where the fifth
force is severely suppressed and thus negligible, but is quite light in low density regions, where the fifth force has a fixed ratio to gravity as in the N models. Furthermore,
as emphasized by \citet{lb2010b}, the fifth force is often not the only impact the coupled scalar field could have on cosmology, nor even is it usually the most important one. 
In the N models, for example, the modification of the cosmic background expansion rate by the scalar coupling could be more influential in the course of structure formation 
\citep{lb2010b}. 

The complexities indicate that the model-independent studies of the coupled scalar field might fail to account for the various effects due to the scalar field coupling
(see \S~\ref{sect:eqns} for a description) appropriately. So in this paper, we have studied the formations and properties of voids in the L, N and C models in parallel and compare their 
predictions.  Voids are the largest objects in the Universe which are produced during the course of structure formation, and fill the vast majority of the space. The importance of 
their properties in understanding the underlying cosmological scenario and global cosmological parameters has been emphasised by many authors. Recently, it has been claimed
that a long-range fifth force could evacuate the space more efficiently and thus produce more voids than $\Lambda$CDM \citep{knp2010, hj2009}. Those studies concentrate on 
the ReBEL model, where only a Yukawa-type fifth force is considered; because of the reason mentioned above, here we take into account all the main effects due to a scalar field coupling (which is arguably the most natural cause of a fifth force), consider two qualitatively different types of models and make more detailed analysis of their effects on void properties by revising the void-finding algorithm.

Our Fig.~\ref{densprob} shows that in the coupled scalar field models matter is more concentrated in some regions, leaving the remaining of the space more evacuated, in line with the expectations. Here we note that the N and C models behave differently: for the former, the fifth force is never suppressed, and the migration of matter from low density regions to high density regions starts earlier thanks to it; for the latter, the chameleon effect suppresses the fifth force at the early times, the effect of which in boosting the clustering of matter only becomes significant recently (after redshift 1). In both models, larger portion of space is under-dense today than in the L model.

We apply our void-finding algorithm to the N and C models, and find that both models predict a bigger number of larger voids than the L model does (Fig.~\ref{vvf}). Once again, the N and C models behave quite differently, in particular at early times: by redshift $1.0$ the N4 model produces several times more voids than the L model (and also the biggest voids are several times bigger), while the C models are only slightly different from the L model, though the fifth force in them, if unsuppressed, is much stronger than in the N models. The result seems to be contradictory to the expectation that in the void regions the fifth force gets less suppressed and therefore we should have seen greater difference from the L model. The reason is as follows: firstly, the scalar field equation of motion is dynamical and the solution in the void regions depends on the overall environment in the simulation box, so it is untrue that the fifth force in void regions is unsuppressed; secondly, the formation rate of voids is more dependent on how fast high density regions could pull matter out of them, but as the fifth force is suppressed this pull is not much stronger than in the L model.

It is worth noting that the difference between the void volume functions in N and L models is bigger at earlier times (Fig.~\ref{vvf}), because at later times most potential void regions have already been developed: since there are not many more new voids yet to be produced in the N models, the L model gradually catches up.  For the C models, the trend is quite opposite, and more voids are produced recently than in the L model, because finally the fifth force is freed and starts to take effect.

We also show in Fig.~\ref{ff} the fraction of space which is filled by voids exceeding a certain threshold in volume. Here the qualitative features could be explained by the same argument used for Fig.~\ref{vvf}, and what is more impressive is the quantitative result it illustrates. For example,  at redshift $1.0$ the N4 model, which is the most extreme in the N models, predicts almost 4 times as much space filled by voids larger than $35~h^{-3}{\rm Mpc}^3$ as does the L model. Even at present the number is almost 2 (the same also applies for the C models). Voids prove to be a promising tool to constrain the scalar field couplings.

Finally, we have studied the density profiles of the voids (Figs.~\ref{vprofile_l}, \ref{vprofile_s}). We find that in general the voids in the coupled scalar field models are featured by a sharper transition from low density to high density around their edges, similar as the result for the ReBEL model \citep{knp2010}. At earlier times the large voids in coupled scalar field models also have lower overdensities in the inner part due to the more effective evacuation of the region.

Due to the limitation of the simulation box and resolution, we have not studied other interesting void properties such as the halos in voids. And because we have only dark matter in the simulations, we have not touched the formations and properties of the void galaxies. These shall be left to future works. The existing results, however, already indicate that the void properties could be largely influenced by a coupled scalar field, being it chameleon or not, and therefore voids could provide a useful tool to study and constrain such alternative scenarios for structure formation. 

\section*{Acknowledgments}

The simulations and processing of data for this work are performed on
the {\tt SARA} supercomputer in the Netherlands, under the {\tt HPC-EUROPA} project, with the support of the European Community Research Infrastructure Action under the {\tt FP8} "Structuring the European Research Area" program, using a
modified version of the publicly available {\tt MLAPM} code \citep{kgb2001}. The author is grateful to Lin Jia for helpful 
discussions and aid in the implementation of the void-finding algorithm. The author
is supported by the Research Fellowship in Applied Mathematics at Queens' College, University of
Cambridge, and the Science and Technology Facility Council ({\tt STFC}) of the
United Kingdom.

\label{lastpage}
\end{document}